\documentclass[twocolumns]{aa}

\usepackage{graphicx}  

\def\ls{{_<\atop^{\sim}}}
\def\gs{{_>\atop^{\sim}}}

\begin{document}

\title{Selection effects shaping the Gamma Ray Burst redshift distributions}

\author{F.~Fiore\inst{1},
D.~Guetta\inst{1},
S.~Piranomonte\inst{1},
V.~D'Elia\inst{1} and
L.A.~Antonelli\inst{1}}

\institute{
INAF --- Osservatorio Astronomico di Roma, via Frascati 33, I-00040
Monteporzio Catone (Roma), Italy
\email{fiore@oa-roma.inaf.it}}

\date{March, 30 2007}

\abstract
{}
{Long Gamma Ray Bursts (GRBs) are associated to the death of massive stars
and have been discovered, so far, up to z=6.29. Therefore, they hold
the promise of probing star-formation and metal enrichment up to very
high redshifts. However, the present GRB samples with redshift
determinations are largely incomplete, and therefore a careful
analysis of selection effects plaguing these samples is mandatory
before any conclusion can be drawn from the observed GRB redshift
distribution.}
{To this purpose we study and compare three well defined samples of long 
GRBs detected by Swift, HETE2 and BeppoSAX.}
{We find that Swift GRBs are, on average, slighly fainter and harder
than BeppoSAX and HETE2 GRBs, as expected due to the higher energy
range (15-150 keV) in which Swift GRBs are detected and localized,
compared to BeppoSAX and HETE2 ($\approx2-20$ keV).

Gas and dust obscuration plays a role in shaping both the GRB samples
and, most interestingly, the present samples of GRBs with redshift
determination. In particular, we argue that the majority of the bright
Swift GRBs without redshift might actually be z$\ls$2 events, and
therefore that the present Swift GRB sample with redshift is biased
against low--z GRBs. On the other hand, the detection of bright UV
rest-frame afterglows from high--z GRBs, and even from those with
large X-ray obscuration, implies a dust amount lower than in nearby
GRBs, and/or a different dust composition. If this is the case, the
Swift sample of GRBs with redshifts is probably a fair sample of the
real high--z GRB population. The absence of high--z GRBs
in the BeppoSAX and HETE2 samples of GRBs with redshifts is probably
due to the fact at the time of BeppoSAX and HETE2 follow-up faint
afterglows of high redshift GRBs will have weaken below the
spectroscopic capabilities of even 10m class telescopes.

The redshift distribution of a subsample of Swift GRBs with
distributions of peak-fluxes, X-ray obscuration and optical magnitude
at a fixed observing time similar to those of the
BeppoSAX and HETE2 samples, is roughly consistent with the real
BeppoSAX+HETE2 redshift distribution.}
{}
%

\authorrunning {Fiore et et al.}
\titlerunning {GRB samples and selection effects }

\maketitle

\section{Introduction}

Gamma Ray Bursts (GRBs) are one of the great wonders of Universe. They
combine several of the hottest topics of 21$^{st}$ century
astrophysics. On one side they are privileged laboratories for
fundamental physics, including relativistic physics, acceleration
processes and radiation mechanisms. On the other side, being some GRBs
associated to the death of massive stars (\cite{mac99}), it was soon
realized after the discovery of their cosmologic origin, that they
could be used as a cosmological tool to investigate star-formation and
metal enrichment at the epochs of galaxy birth, formation and growth
(e.g. Wijers et al. 1998, Porciani \& Madau 2001, Fynbo et al. 2007).
In this respect, two main research areas have developed.  The first
one uses GRBs as background beacons for spectroscopy of UV lines to
characterize the physical and chemical state of the matter along the
line of sight (Savaglio 2006 and references therein).  The second
includes statistical studies of the GRB redshift distributions
(\cite{guetta05}, \cite{nata05}, \cite{jako06}, \cite{daigne06}).
Even though the techniques adopted, and therefore the reference
communities, are somewhat different, these research areas are
interconnected. As an example, UV lines can be used to determine the
metal content of the absorption systems (\cite{delia06},
\cite{delia06b}, \cite{procha06}, Savaglio 2006).  On the other hand,
it is well known that galaxy-scale properties like metalicity,
star-formation rate and mass are correlated. Then, metallicity
determinations obtained through GRB spectroscopy can, at least in
principle, be plugged in the GRB population studies, to obtain a
better constraint on the models.  In this paper we concentrate on
population studies and in particular on the importance of selection
effects in shaping GRB redshift distributions.

\begin{figure*}
\begin{tabular}{cc}
\includegraphics[height=8.5truecm,width=8.5truecm]{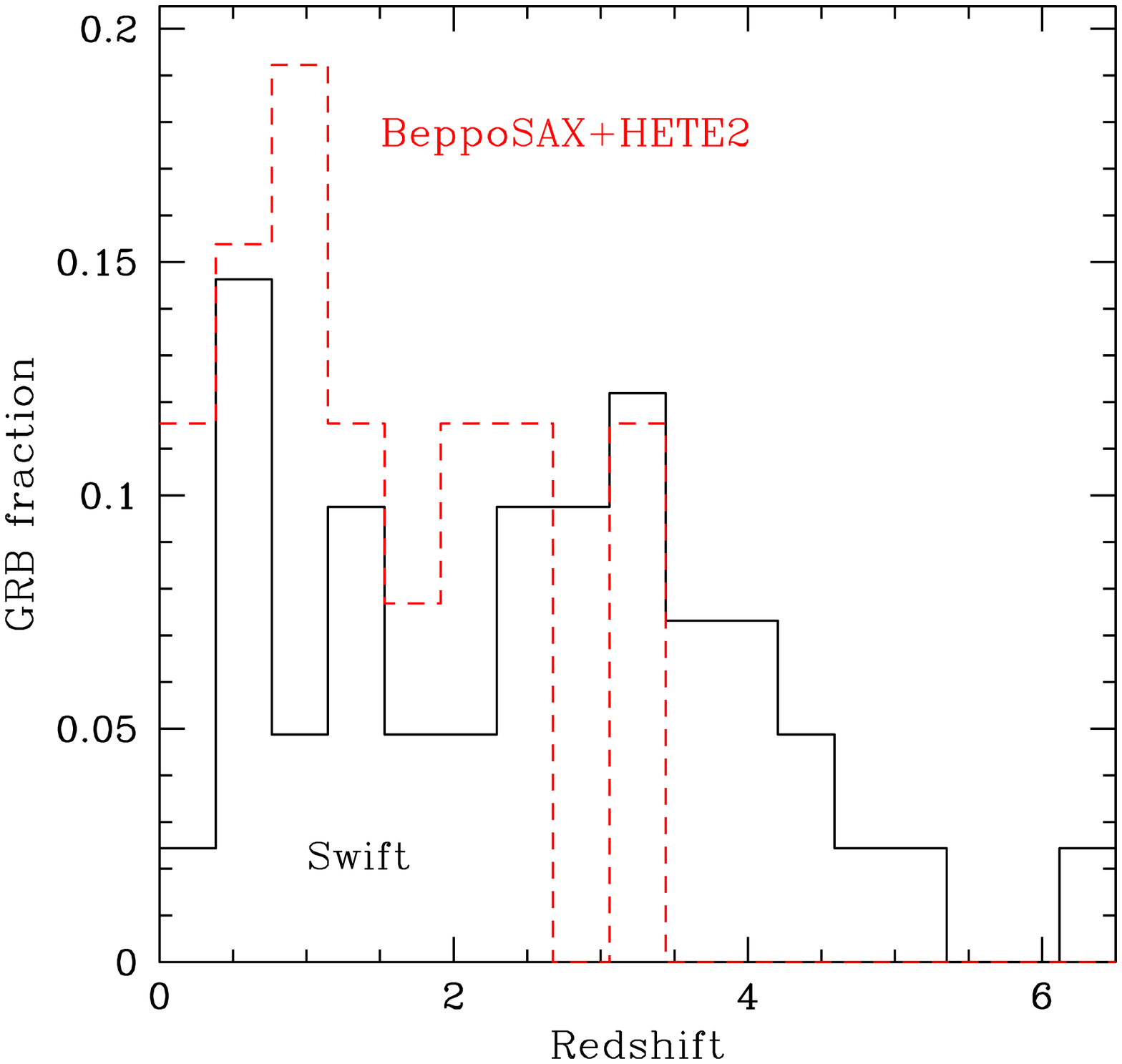}
\includegraphics[height=8.5truecm,width=8.5truecm]{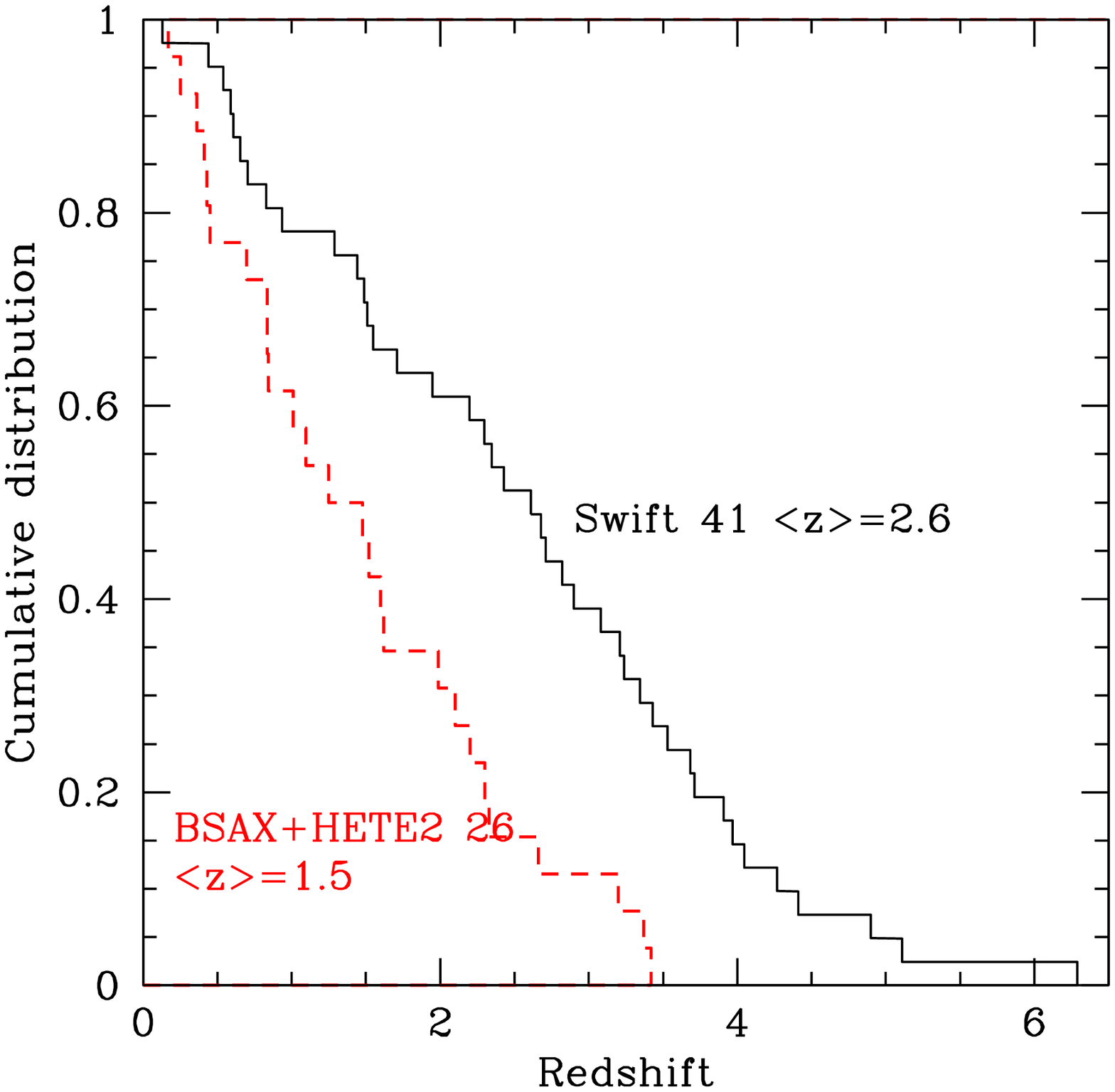}
\end{tabular}
\caption{Redshift distribution (left panel) and cumulative redshift
distributions of Swift (solid line) and BeppoSAX+HETE2 (dashed line)
GRBs.}
\label{zdist}
\end{figure*}

Population studies are very powerful tools. For example, galaxy and
AGN counts and luminosity functions have been used to successfully
measure the evolution of the star-formation rate, galaxy and black
hole mass densities up to z$\ls6$. Similarily, GRBs can be used to
probe the histories of the GRB- and star- formation rates and of the
metal enrichment in the Universe (e.g. \cite{pm01}). Indeed, thanks to
BeppoSAX first and then to HETE2 and Swift we begin having sizable
samples of GRBs with reliable redshifts (about 80 up to now). This
number should grow up to 150-200 within the Swift lifetime.  This
opens up the possibility to compute fairly well constrained GRB
luminosity functions in a few redshift bins, and therefore measure the
cosmic evolution of the GRB rate. The fraction of Swift GRBs
with a reliable redshift is today about one third of the total. It might be
expected that this fraction will improve in future, but it will hardly
approach the majority of the GRBs. This means that the biggest problem
we have to face in exploiting GRBs as cosmological tools is to understand and
account for large selection effects.  The role of large
selection effects in shaping the population of GRBs with a measured
redshift is evident when comparing the redshift distribution of Swift
GRBs with that of BeppoSAX and HETE2 GRBs (figure \ref{zdist}). The
median redshifts of the two distributions are 2.6 and 1.5
respectively.  This discrepancy cannot be explained simply as due to
the different detector sensitivity (e.g. Guetta \& Piran 2007).

In the next sections we make a detailed description of what are
the possible selection effects that plague the GRB redshift determination.

\section{Samples used in this study}

To gain more quantitavive information on the issue of GRB selection
effects we study three well defined sample of GRBs detected by Swift,
HETE2 and BeppoSAX. We select long GRBs (T90$>$3 sec) outside the
Galactic plane to limit Galactic extinction along their line of sight
and to avoid too crowded fields, which can complicate the discovery of
optical/NIR afterglows, and thus hamper redshift determinations. To
this purpose we limit our study to regions with Galactic column
density along the line of sight smaller than $\times10^{21}$ cm$^{-2}$
(corresponding to A$_V\ls1$).  We also select GRBs with good (arcmin)
localization.  For BeppoSAX and HETE2 GRBs we require that the
$\gamma-$ray burst is detected by the high energy GRBM and FREGATE
instruments and is localized by the WFC, WXC or SXC instruments. For
Swift we consider all long GRBs detected before September 10 2006,
while for HETE2 we consider all long GRBs detected up to December 31
2003 . For BeppoSAX we consider all GRBs detected during the entire
mission. We excluded from the sample GRB 060218 and GRB980425, which
are probably associated to a different class of events, orders of
magnitude fainter than the rest of the sample (e.g.  Guetta \& Della
Valle 2007).  We consider only reliable spectroscopic redshifts. Table
1 gives more information on the selected samples.

\begin{table*}
\caption{\bf GRB samples}
\begin{tabular}{lccccc}
\hline
Sat.      & Tot. GRB & O.A.$^a$ & O. decay$^b$ & Tot. z spec.$^c$ & 
z from em. lines$^d$ \\
\hline
Swift     & 122      & 62 & 44    &   41        &  6 \\
BeppoSAX  & 39       & 18 & 16    &   12        &  3 \\
HETE2	  & 44       & 17 & 15    &   14        &  2 \\
\hline
\end{tabular}

$^a$ GRB with Optical Afterglows; $^b$ GRB with multiple optical
observations and estimated optical afterglow temporal decay index;
$^c$ total number of GRBs with a reliable spectroscopic redshift;
$^d$ GRBs with a redshift derived only through spectroscopy 
of the host galaxy.

\end{table*}

Swift BAT peak-fluxes and spectral parameters are taken from the {\it
Swift GRB Information page}\footnote{
http://swift.gsfc.nasa.gov/docs/swift/archive/grb\_table.html}. 

Equivalent hydrogen column densities (N$_H$) are computed from X-ray
afterglow spectra assuming solar abundances. Swift column densities
and are taken from (\cite{cam06}) in 17 cases, from our own analysis
in 12 cases and from the {\it Swift GRB Information page} in the rest
of the cases.  BeppoSAX peak-fluxes and spectral parameters, including
hydrogen equivalent column densities, are taken from Stratta et
al. (2004), Piro et al.  (2005), and De Pasquale et al. (2006).  For
both samples the minimum column density is set to the Galactic value
along the line of sight (Dickey \& Lockman 1990).  HETE2 peak-fluxes
and spectral parameters are taken from Sakamoto et al. (2005).

Swift optical afterglow parameters are taken from the GCN throght the
{\it Gamma Ray Burst
database}\footnote{http://grad40.as.utexas.edu/grblog.php}. BeppoSAX
and HETE2 optical afterglow parameters are taken from the {\it Gamma
Ray Burst database}, from Zeh et al. (2006) and from De Pasquale et
al. (2006).  Table 1 gives the number of optical afterglow detections
and the number of cases in which multiple observations allowed us to
estimate the optical afterglow decay index. We use wherever possible R
band magnitudes. For 17 Swift GRBs and 1 BeppoSAX GRB we have only V
band magnitudes, for 2 Swift GRBs only a white filter magnitude and
for 8 BeppoSAX GRBs only g-band magnitudes. For all these GRBs
we converted the observed magnitude into the R band using standard
afterglow colors.  The Lyman$-\alpha$ forest starts to enter the R
band at z=3.9. Therefore, R band magnitudes for the GRBs at z=4-5
should be considered lower limits. For GRB050904 at z=6.29 we used
TAROT I band equivalent magnitudes (Boer et al. 2006).

Table 1 gives also the number of reliable spectroscopic redshift
obtained for the three samples.  In most of the cases the redshift has
been obtained through absorption lines overimposed on the afterglow
spectrum. In a minority of cases (6, 3 and 2 for the Swift, BeppoSAX
and HETE2 samples respectively) the redshift has been obtained
uniquely through spectroscopy of the host galaxy, being the optical
afterglow undetected or too faint to search for absorption
features. In a few other cases the redshift has been obtained thanks to
both absorption lines in the optical afterglow emission
and host galaxy emission lines.

There are at least two large groups of selection effects that must be
considered: (1) GRB detection and localization and (2)
redshift determination through spectroscopy of the optical/NIR
afterglow or of the GRB host galaxy. We discuss these two issues in
the next sections.

\section {GRB detection and localization}

The sensitivity of BeppoSAX, HETE2 and Swift instruments as a function
of the GRB spectral shape has been studied in detail by Band (2003,
2006).  Band (2006) also studied the sensitivity of the BAT instrument
as a function of the combined GRB temporal and spectral properties. We
refer to these papers for more details on these topics.

\begin{figure*}
\begin{tabular}{cc}
\includegraphics[height=8.5truecm,width=8.5truecm]{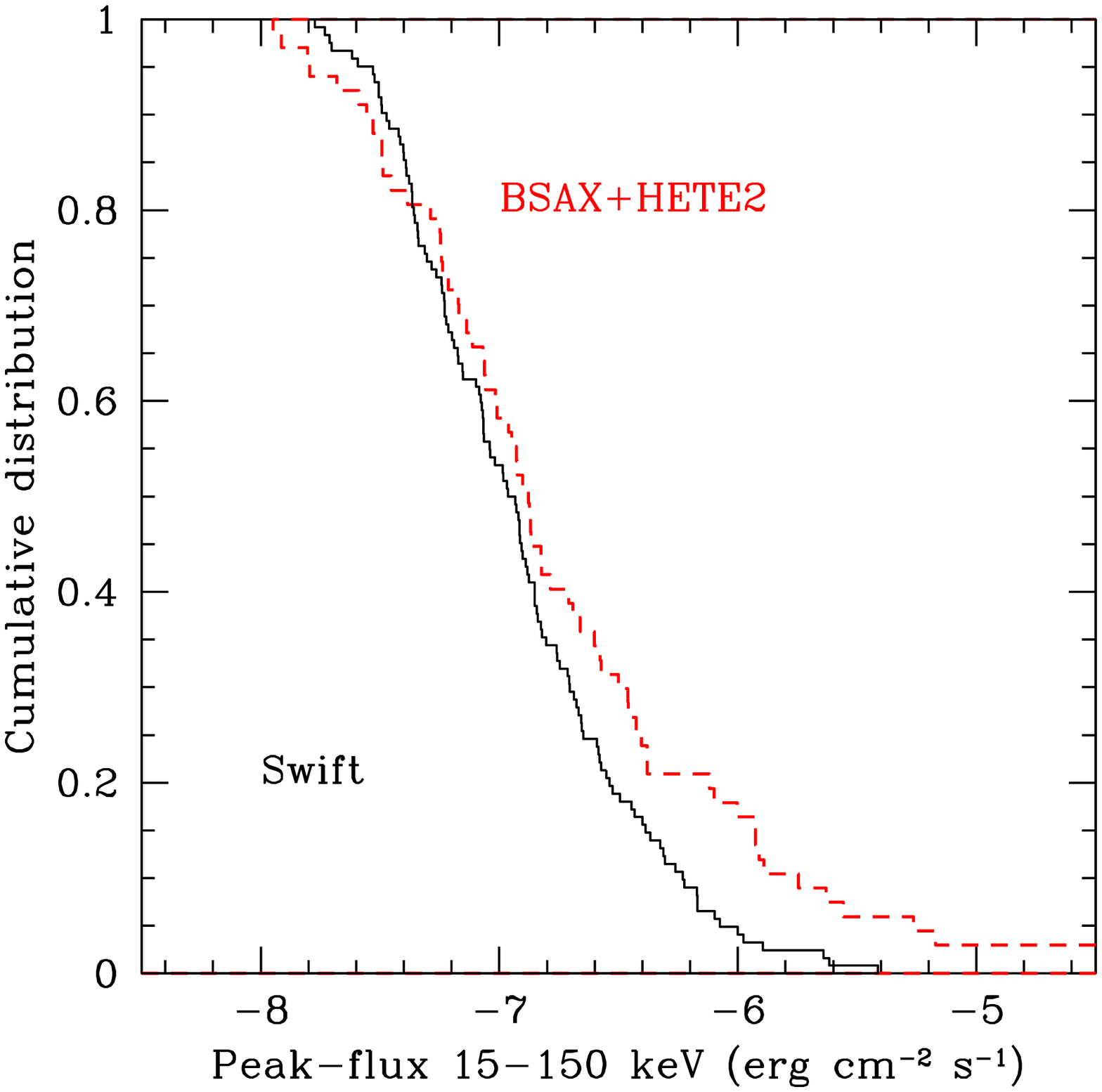}
\includegraphics[height=8.5truecm,width=8.5truecm]{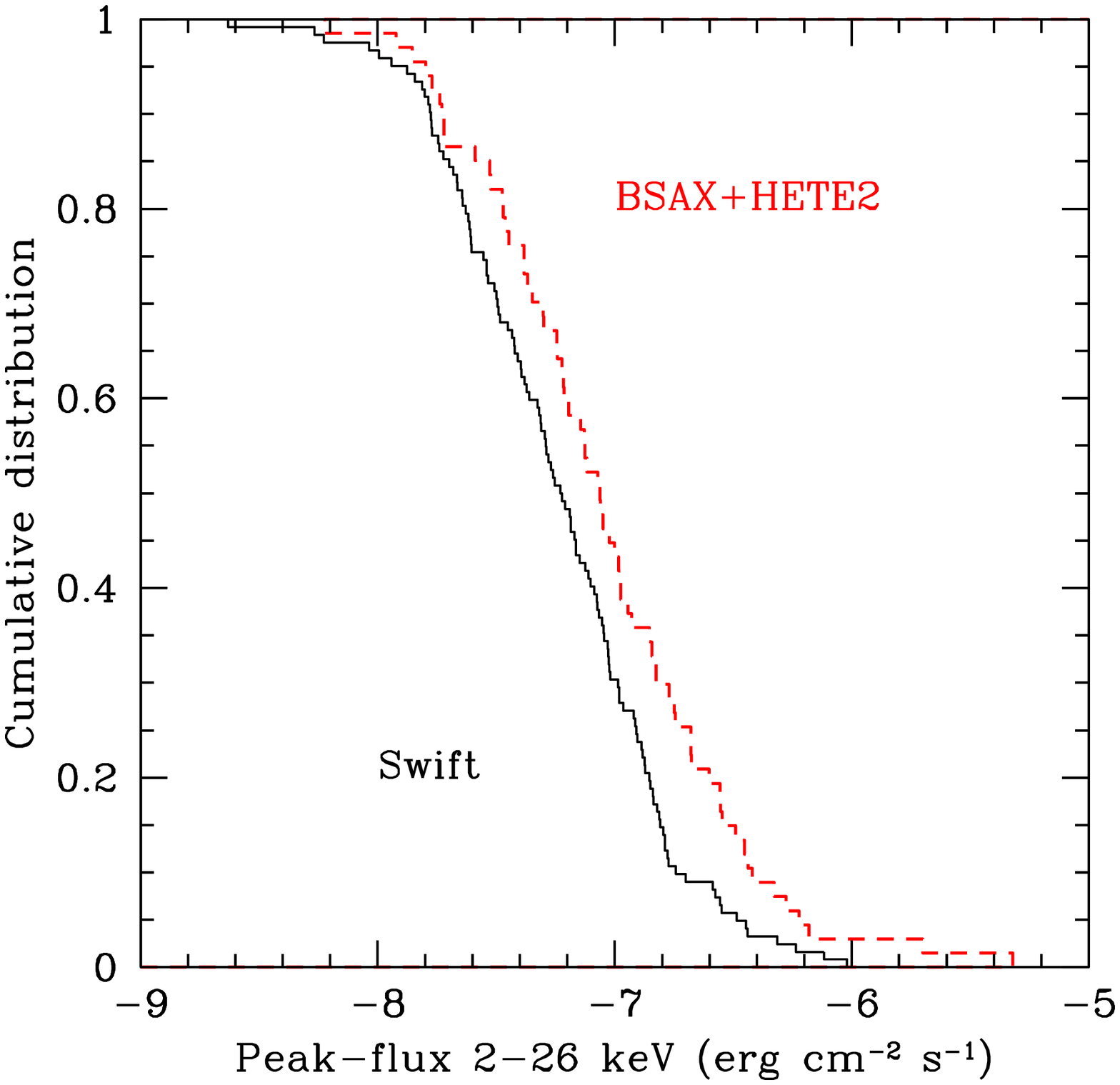}
\end{tabular}
\caption{Peak-flux cumulative distributions of the Swift (solid line) and
BeppoSAX+HETE2 (dashed line) GRBs.  a), left panel, 15-150 keV band; b), right
panel, 2-26 keV band.}
\label{pfdist}
\end{figure*}

Figure \ref{pfdist} compares the peak-flux cumulative distributions of
the Swift GRBs with that of BeppoSAX and HETE2. The comparison is done
in two energy bands: 15-150 keV, which is the band where BAT detects
and localizes GRBs, and 2-26 keV, which is the band where the BeppoSAX
WFC and the HETE2 WXC and SXC localize GRBs. To produce figure
\ref{pfdist}a) BeppoSAX GRBM and HETE2 Fregate peak-fluxes were
converted to the 15-150 keV BAT band by using a power law model with an
(average) energy index of 0.5 for the BeppoSAX burts and the best fit
model in Sakamoto et al. (2005) for the HETE2 bursts.  To produce
figure \ref{pfdist}b) we used WFC and WXC peak-fluxes and converted
BAT 15-150 keV peak fluxes in the 2-26 keV band by using the best fit
models and parameters and the best fit observed column densities along
the line of sight to the GRBs. To assess the robustness of our
analysis we produced peak-flux cumulative distributions using
different, but reasonable, values of the spectral parameters adopted
for the conversion from one band to the other. We always found 
qualitatively similar results to those in figure \ref{pfdist}.

Figure \ref{pfdist}a) shows that Swift finds, on average, 
slightly fainter GRBs than BeppoSAX and HETE2 in the 15-150 keV band. The
BeppoSAX and HETE2 samples contain a higher fraction of bright
GRBs. The median log(peak-flux) and its interquartile range are -6.93,
0.33 for the Swift sample and -6.88, 0.42 for the joined
BeppoSAX+HETE2 sample. This is expected because of the better
sensitivity of the BAT instrument with respect to the BeppoSAX GRBM
and HETE2 Fregate instruments (\cite{b03}).

The median 2-26 keV log(peak-flux) is -7.22, 0.32 for the Swift sample
and -7.06, 0.38 for the joined BeppoSAX+HETE2 sample. The two 2-26 keV
peak-flux distributions differ from each other more than the 15-150
keV distributions.  This is probably due to the fact that Swift GRBs
are localized at energies higher than 10-15 keV, while BeppoSAX and
HETE2 GRBs are localized at energies $\ls10$ keV.  This implies that
Swift localizes, on average, harder GRBs than BeppoSAX and
HETE2. In particular, Swift GRBs are revealed in a spectral range in
which absorption has little, if any, effect.
A column density of N$_H=10^{23}$ cm$^{-2}$
at z=1 would reduce the observed 2-10 keV flux by 12-15\% (depending
on the spectral index), thus reducing the probability of detecting
such a highly obscured GRBs with BeppoSAX WFC and HETE2 WXC. Conversely,
these GRBs would certainly be present in the Swift sample.

\begin{figure*}
\begin{tabular}{cc}
\includegraphics[height=8.5truecm,width=8.5truecm]{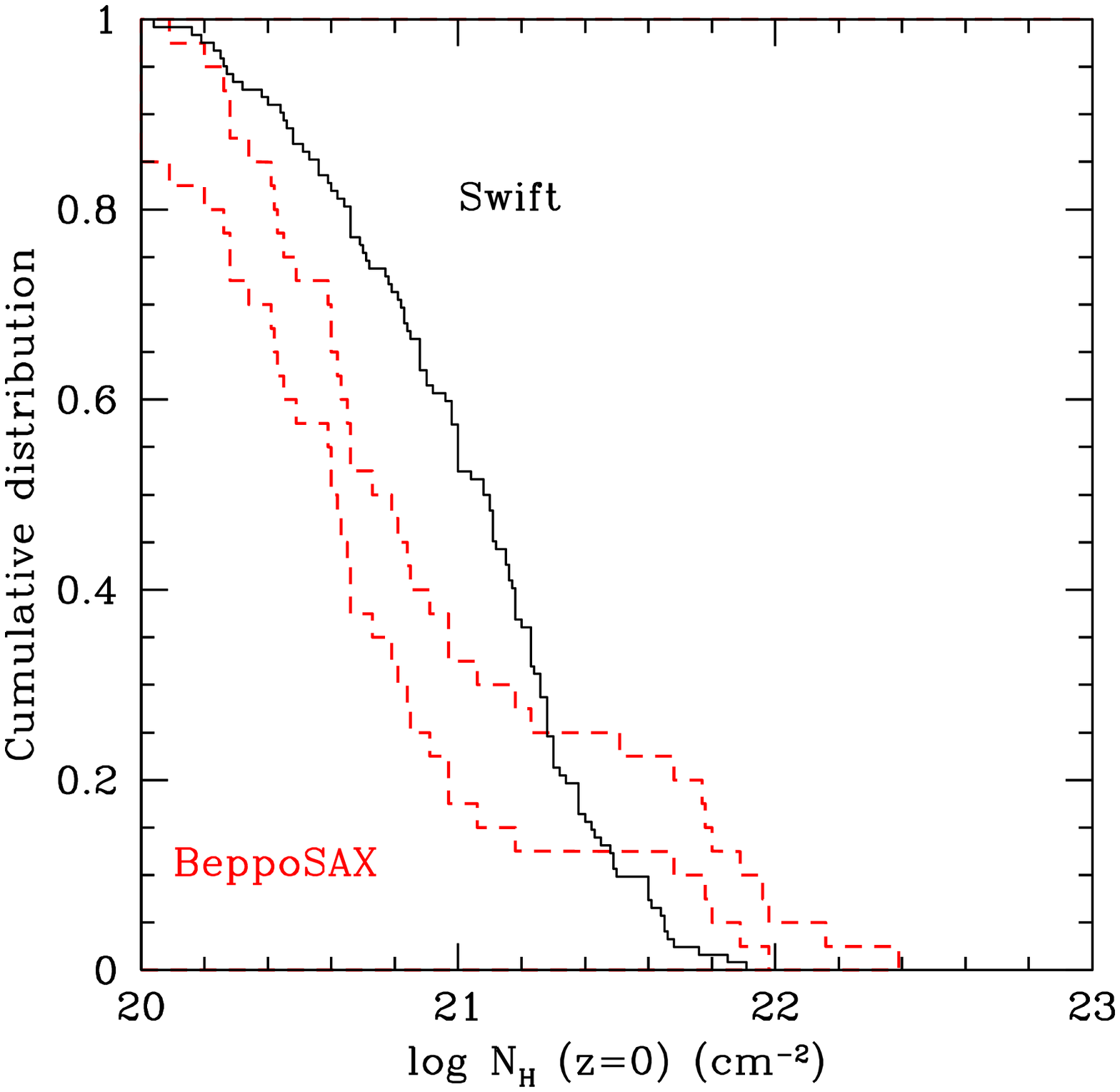}
\includegraphics[height=8.5truecm,width=8.5truecm]{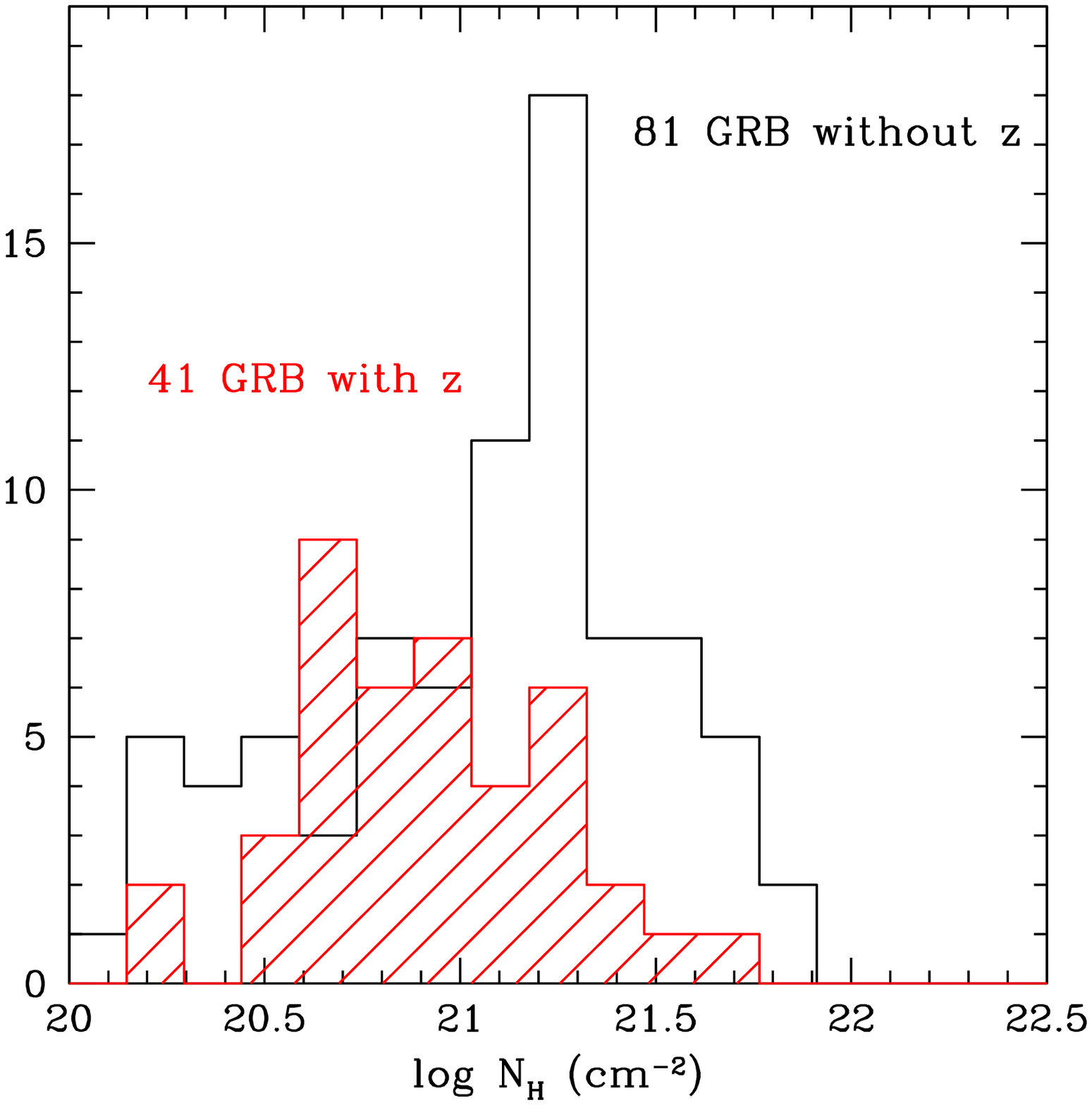}
\end{tabular}
\caption{a), left panel, N$_H$ cumulative distributions of the Swift
(solid line) and BeppoSAX (dashed lines) GRBs. The leftmost BeppoSAX
curve assumes N$_H$=N$_H{Galactic}$ for the GRBs with a best fit
intrinsic N$_H$ consistent with zero. The rightmost BeppoSAX curve
assumes for these GRBs the 90\% upper limit. b), right panel, N$_H$
histograms of the Swift GRB with (shadow histogram) and without
redshift (black histogram).}
\label{nhdist}
\end{figure*}

Figure \ref{nhdist}a) compares the best fit column density N$_H$ in
observer frame for the samples of Swift and BeppoSAX GRBs.  The X-ray
afterglows at the time of the BeppoSAX NFI observations (obtained
repointing the satellite with a typical delay time of 8-10 hours from
the GRB event), were significantly weaker than at the time of the
Swift observations (typically minutes to a few hours after the GRB
event), due to the afterglow power law decrease with exponent
$\gamma=-1:-2$.  This implies that the uncertainties on the X-ray
spectral parameters, and therefore N$_H$, are much bigger for BeppoSAX
GRBs than for Swift GRBs. Indeed, the typical uncertainty of Swift
column densities is $5-10\times10^{20}$ cm$^{-2}$ (see e.g. Campana et
al. 2006), whereas that of BeppoSAX is $\approx10$ times larger (see
Stratta et al. 2004 and De Pasquale et al. 2006).  For this reason we
plot 2 curves for the BeppoSAX GRBs. The leftmost curve assumes
N$_H$=N$_H(Galactic)$ for those GRBs whose best fit intrinsic N$_H$ is
consistent with zero. The rightmost curve is based on 90\% upper
limits on the N$_H$ of these GRBs. The tail at high  N$_H$ values of 
this distribution is due to not well constrained upper limits.
The real BeppoSAX N$_H$ distribution is probabily between the two curves. 

Shortward of a few$\times10^{21}$ cm$^{-2}$ the BeppoSAX curves in
figure \ref{nhdist}a) are significantly lower than the Swift
curve. The probability that the BeppoSAX and Swift curves are drawn
from the same parent population is $<10^{-5}$ and 1.7 \% respectively,
using the Kolmogorov-Smirnov test, thus confirming that Swift samples
are less biased against obscuration than the BeppoSAX sample. Since
the observer frame column density scales as the rest frame column
density times (1+z) to a large negative power ($\sim-2.5$), this
implies that the BeppoSAX sample is somewhat biased against low-z,
highly obscured GRBs. Conversely, these GRBs must be present in the
Swift sample.

Figure \ref{nhdist}b) compares the N$_H$ distribution of the Swift
GRBs with determined redshift to that of the Swift GRBs with
undetermined redshift. The probability that the two distributions are
drawn from the same parent population is only 1\%, suggesting that the
sample of Swift GRBs with determined redshift is biased against GRB
with large (observer frame) obscuration. Indeed, the N$_H$
distributions of the Swift and BeppoSAX GRBs with redshifts are
similar, unlike the N$_H$ distributions of the full Swift and BeppoSAX
GRBs, see above.  This introduces the next important group of
selection effects, those related to the determination of the redshift
of a GRB through spectroscopy of the optical/NIR afterglow or of its
host galaxy.

\section{Redshift determination}

In determining the redshift of a GRB the identification of the optical
afterglow plays a major role. Only 6 Swift redshifts have been found
through spectroscopy of the host galaxy (5 for the BeppoSAX and HETE2
joined sample).

Optical afterglows have been discovered for only 50\% of the Swift GRB
sample, a fraction only slightly greater than that of the BeppoSAX and
HETE2 samples (46\% and 39\% respectively).  This result is somewhat
surprising, in considetation of the prompt Swift localization
(minutes) and the large international effort on Swift GRB follow-up
observations, which exploits an impressive number of facilities, from
dedicated robotic telescopes to 8m class telescopes like the VLT,
Gemini and Keck. It was expected that such an effort would have
produced a much larger fraction of optical/NIR afterglow
identifications than BeppoSAX and HETE2.

\begin{figure}
\includegraphics[height=8.5truecm,width=8.5truecm]{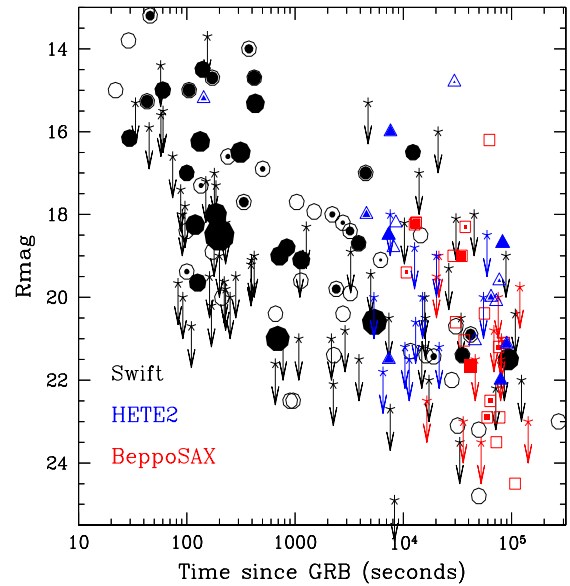}
\caption{The R magnitude of the optical afterglow at the time of its
discovery as a function of this time. Filled symbols are GRBs with
reliable redshift determination. The size of the symbol is
proportional to the redshift (the larger the symbol the larger the
redshift).  Circles = Swift GRBs; squares = BeppoSAX GRBs; triangles =
HETE2 GRBs.}
\label{r3}
\end{figure}

Figure \ref{r3} shows the R magnitude of the optical afterglow as a
function of the time of discovery of the optical afterglow the Swift,
BeppoSAX and HETE2 GRBs. As expected, redshifts are preferentially
found for bright afterglows.  The figure suggests also that at a given
time from the GRB event the Swift optical afterglows are fainter, on
average, than the BeppoSAX and HETE2 afterglows. We then computed the
magnitude of the Swift, BeppoSAX and HETE2 afterglows at a fixed time
using the best fit decay indices found for each GRB afterglow, when
available. In the rest of the cases we used a time decay index of
-1. We chose a fixed time of 10ks after the burst (observer frame),
which is intermediate between the typical times at which Swift,
BeppoSAX and HETE2 GRBs are discovered, thus minimizing the
extrapolation to compute the R mag at 10ks.

\begin{figure*}[t]
\begin{tabular}{cc}
\includegraphics[height=8.5truecm,width=8.5truecm]{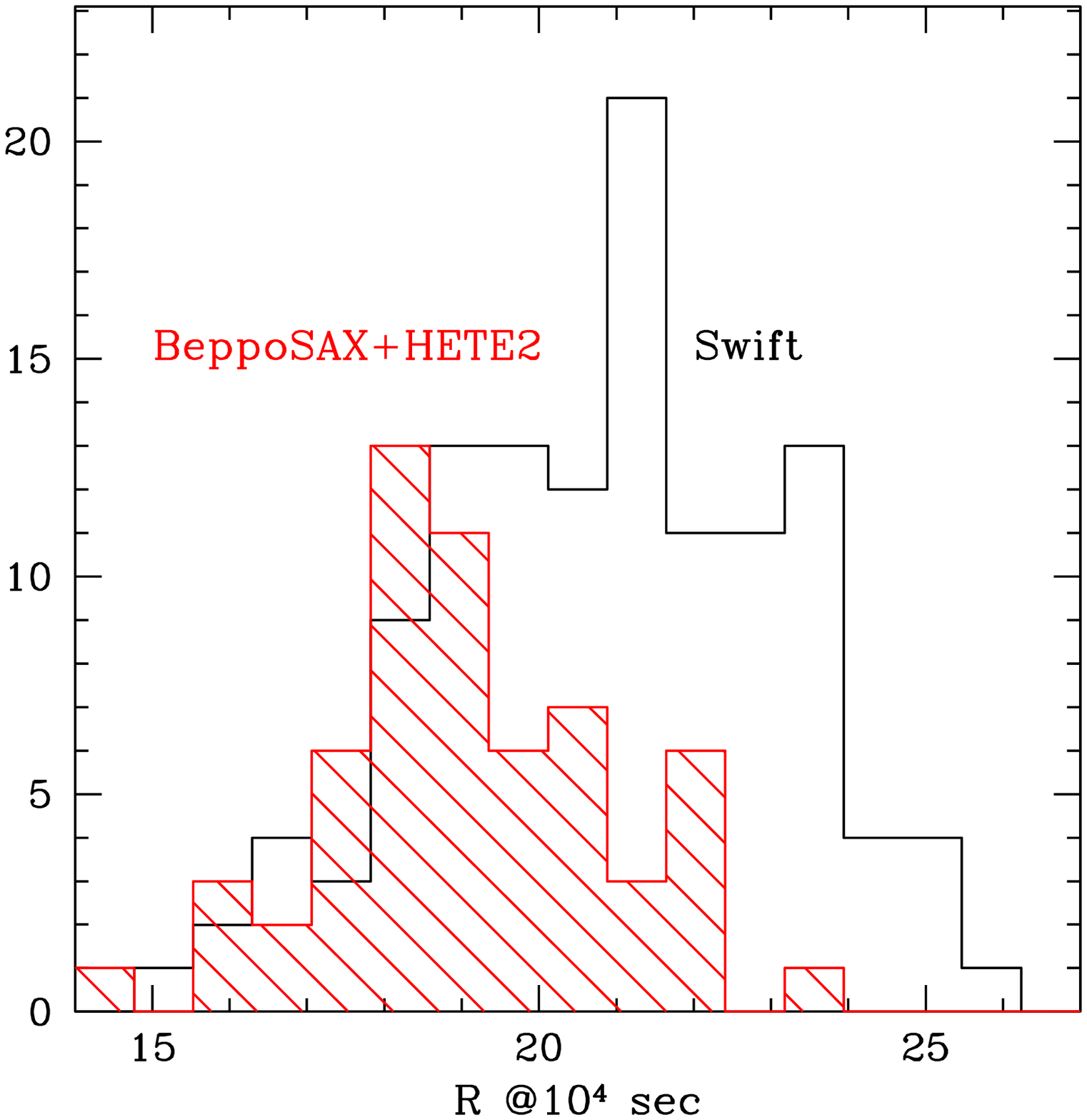}
\includegraphics[height=8.5truecm,width=8.5truecm]{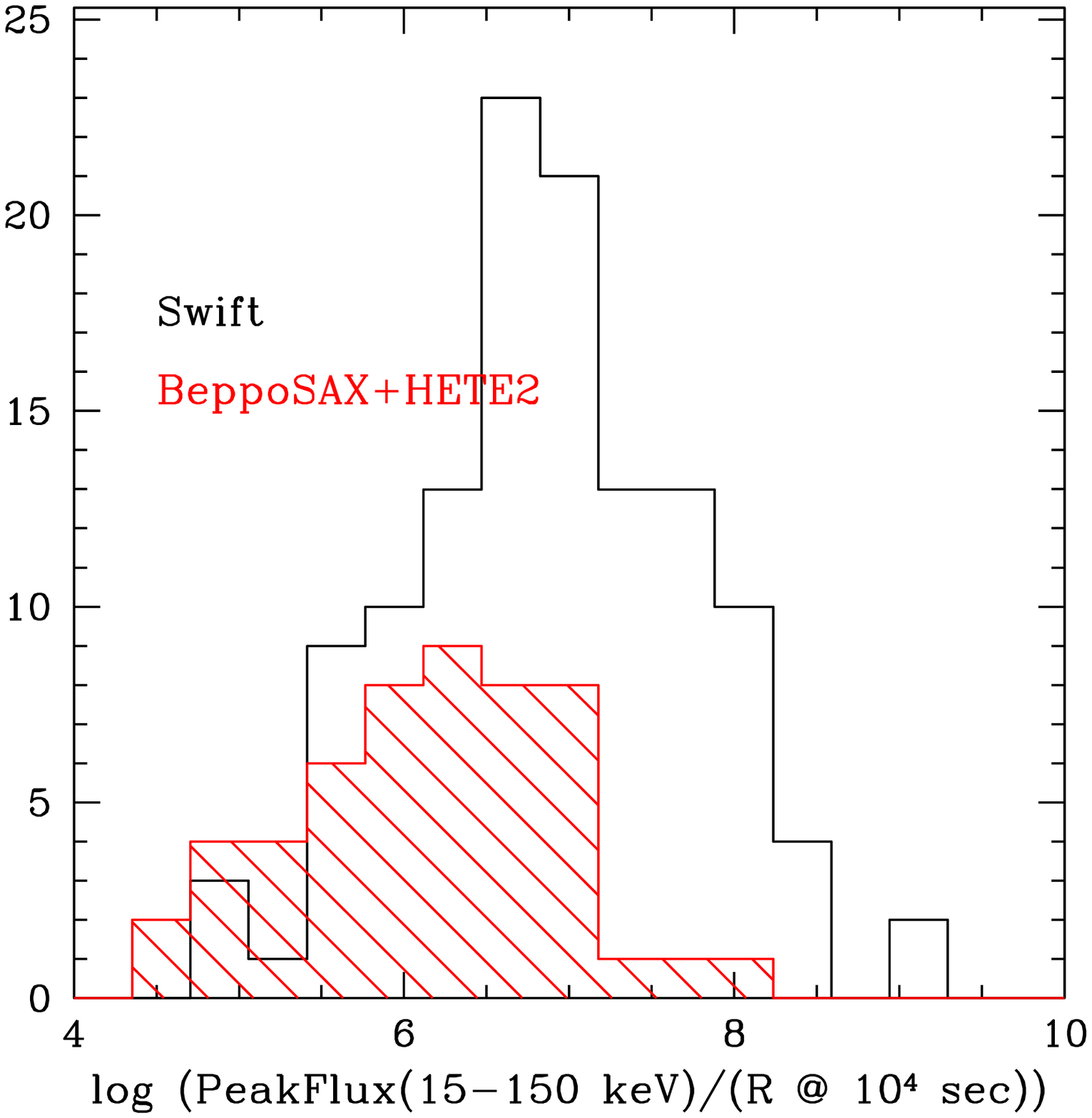}
\end{tabular}
\caption{a), left panel, the distribution of the R magnitude 10ks
after the GRB events for the Swift (solid histogram) and the
BeppoSAX+HETE2 (dashed histogram) GRB samples. GRBs without detection
of optical afterglow but for which optical follow-up observations were
carried out are included in this distribution at the magnitude of the
corresponding upper limits. b), right panel, the distribution of the
$\gamma-$ray (15-150 keV) to optical (R band) flux ratio for the Swift
(solid histogram), and BeppoSAX+HETE2 (red histogram) GRB samples.}
\label{lxo}
\end{figure*}

Figure \ref{lxo}a) compares the Swift distribution of the R mag at 10
ks from the GRB event with that of the BeppoSAX and HETE2. GRBs
without detection of optical afterglow but for which optical follow-up
observations were carried out are included in this distribution at the
magnitude of their upper limits.  This figure confirms that Swift
finds bursts with a fainter optical afterglow. The probability that
the Swift and BeppoSAX+HETE2 distributions are drawn from the same
parent population is $<10^{-5}$. Similar results are obtained by
considering the distributions of the magnitudes of the detected
afterglows, excluding the upper limits. In principle, the fainter
Swift optical afterglows may be due to the fact that Swift detects, on
average, fainter GRBs (see figure \ref{pfdist}). However, this is
probably not the case. Figure \ref{lxo}b) shows the Swift and
BeppoSAX+HETE2 distributions of the $\gamma-$ray (15-150 keV) to
optical (R band) flux ratio. (Also in this case GRBs with undetected
optical afterglow are included at the magnitude of their upper
limits.) The probability that the two distributions are drawm from the
same parent population is smaller than 1\%. This probability increases
to 1.6\% by comparing the distributions of the magnitudes of the
detected afterglows, excluding the upper limits.  Similar results are
obtained considering the X-ray (2-26 keV) to optical flux ratio.
Computing the R magnitude at 1 ks or at 100 ks does not change
qualitatively this result.

\section{Selection effects at work}

There two major differences in the Swift and BeppoSAX+HETE2 redshift
distribution: a) a relatively large number of GRB with z$>$3.5 is
present in the Swift sample (11 out 41 GRB, i.e. 27 \% of the
sample). These GRBs are absent in the combined
BeppoSAX+HETE2 sample. b) a deficit of low redshift (z$\ls2$) in the
Swift sample with respect to what would be expected based on the
BeppoSAX+HETE2 sample. We discuss these two points in turn.

\begin{figure*}[th!]
\begin{tabular}{cc}
\includegraphics[height=8.5truecm,width=8.5truecm]{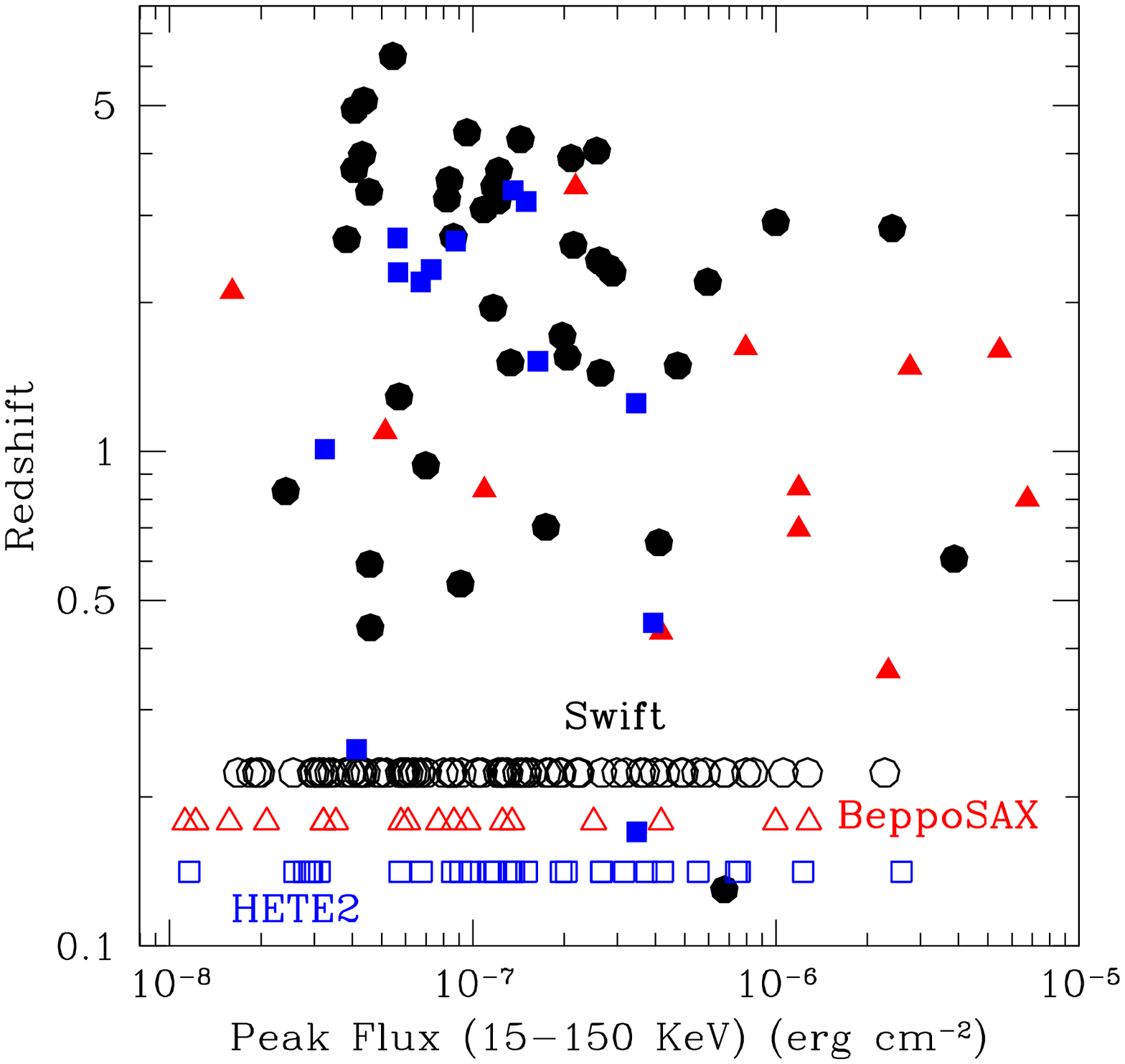}
\includegraphics[height=8.5truecm,width=8.5truecm]{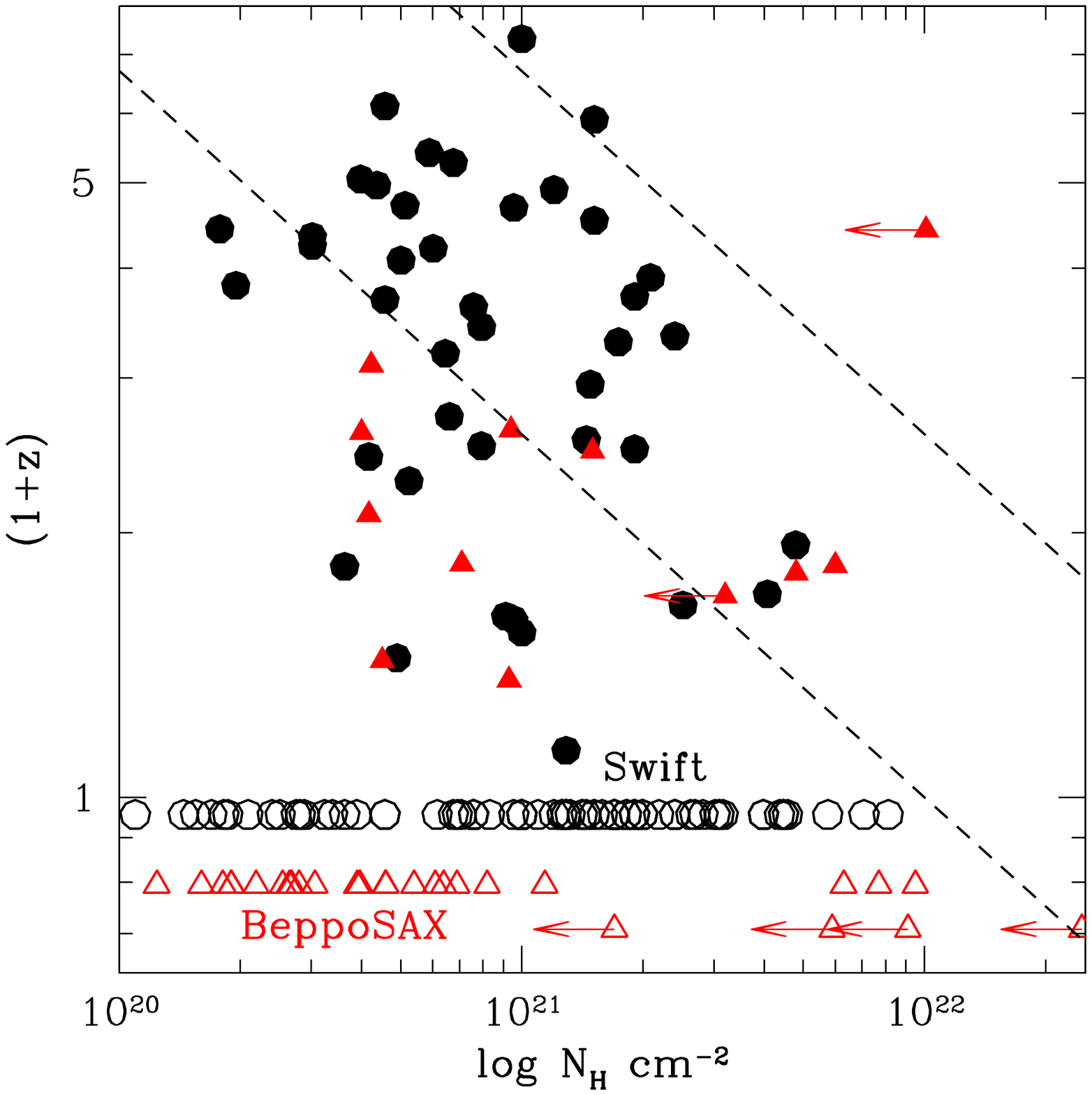}
\end{tabular}
\caption{The redshift as a function of the 15-150 keV peak-flux (a),
left panel) and of the observer frame N$_H$ (b), right panel) for the
Swift (cycles), BeppoSAX (triangles) and HETE2 (squares) GRBs. GRB
without a measured redshift are plotted at a constant z as empty
symbols.  The left dashed line in figure \ref{nhz}b) is the
expectation for a constant, rest-frame column density of log$N_H=22$,
the right dashed line is the expectation for log$N_H=23$.}
\label{nhz}
\end{figure*}

About the first point, the Swift better sensitivity to faint GRBs and
the Swift quick localization may explain the presence of a large
number of high redshift GRBs in the Swift sample compared to the
BeppoSAX and HETE2 samples. First, the highest redshift GRBs are found
at low peak-fluxes in figure \ref{nhz}a), which plots the redshift as
a function of the 15-150 keV peakflux for the Swift, BeppoSAX, and
HETE2 GRB samples. Second, the Swift capability to localize the GRB on
time-scales of minutes allows the discovery of faint optical
afterglows, that can be promptly observed in spectroscopic mode. The
median delay time of optical follow-up for the Swift, HETE2 and
BeppoSAX GRBs is 15 minutes, 3.5 hours and 14 hours respectively.  If
optical and near infrared afterglows decreases like power laws with
exponent $\gamma\approx-1$, they would have faded by 2.9 and 4.4
magnitudes passing from the median Swift delay time to the median
HETE2 and BeppoSAX delay times, respectively.  Faint afterglows of
high redshift GRBs will have weakened even below the spectroscopic
capability of 10m class telescopes, if observed many hours later
like in the BeppoSAX and HETE2 era. Furthermore, the host galaxies of
high redshift GRBs are too faint to allow redshift determinations
through their emission lines.

The discrepacy between the Swift and BeppoSAX+HETE2 samples at low
redshift is less straightforward and requires a more detailed
discussion.  Figure \ref{nhz}a) shows that the peak-flux
distrubution becomes wider at low redshift. Indeed, the median
redshift of the 24 Swift GRB with 15-150 keV peak-flux
$>3\times10^{-7}$ erg cm$^{-2}$ ($\sim20\%$ of the sample) is only
$<$z$>$=1.5, very different from the median redshift of the full
sample ($<$z$>$=2.6).  
The sample of bright GRBs is particularly useful because: a)
selection effects due to temporal and spatial variation of the
instrument sensitivity are minimized; and b) the redshift range is
narrower, being high--z GRBs systematically fainter than bright GRBs,
thus minimizing evolutionary effects. 
For bright fluxes the sensitivity of the
instruments can be safely considered constant over their entire field
of view, and it is therefore easier to compare the number of GRBs
expected by different experiments.  Comparing the field of view of
Swift BAT to that of the BeppoSAX WFC, and considering the net
observing time spent by the two satellites searching for GRBs, we
expect a number of bright GRBs (15-150 keV peak-flux $>3\times10^{-7}$
erg cm$^{-2}$) $\sim1.5$ times higher in the Swift sample than in the
BeppoSAX sample, a factor similar to that found in the real GRB
samples (1.77). Conversely, the number of bright GRBs with z$<$2 in
the Swift sample is only half that in the BeppoSAX sample (4 against
8).  It is clear that a strong selection effect is at work, biasing
the sample of Swift GRBs with redshift against low--z sources. Indeed,
only 7 out 24 bright Swift GRBs have a spectroscopic redshift, to be
compared to 8 out 13 in the BeppoSAX sample (and 3 out 6 of the HETE2
sample).

A possible cause of the difficulty in obtaining a redshift for many
bright Swift GRBs is obscuration. The median observer-frame column
density toward the bright Swift GRBs is log$N_H=21.28$ with an
interquartile range of 0.24, while the median logN$_H$ of the faint
Swift GRBs is logN$_H=21.0$ with interquartile 0.30.  The probability
that the two logN$_H$ distribution are drawn from the same parent
population is $\ls2\%$. The median log$N_H$ of the 13 BeppoSAX GRBs
with 15-150 keV peak-flux $>3\times10^{-7}$ erg cm$^{-2}$ is 20.66 (or
20.97 assuming the 90\% upper limit for the GRBs with a best fit
intrinsic N$_H$ consistent with zero). The intrinsic observer-frame
logN$_H$ (i.e. after subtraction of the Galactic column density along
the line of sight) of the 24 Swift bright GRBs is 21.10.  At a typical
redshift of 1.5 this implies a rest-frame column density of
logN$_H\sim22.1$ and an optical extinction of several magnitudes,
assuming a Galactic dust to gas ratio. This dust extinction would make
more difficult both the discovery of optical afterglows and the
determination of the redshift through optical spectroscopy.  Indeed,
the fraction of detected optical afterglows among the bright Swift
GRBs is 46\%, slightly smaller than that of the sample of the 98 Swift
GRBs with 15-150 keV peak-flux $<3\times10^{-7}$ erg cm$^{-2}$ (52
\%). The median R band magnitudes of the bright GRBs ($<$R$>$=18.4) is
also similar to that of the faint GRBs ($<$R$>$=18.7). Nearly
identical are the median R magnitudes at 10 ks, $<$R$>$20.73 for
bright GRBs and $<$R$>$20.72 for the faint GRBs. Conversely, one would
expect fainter optical afterglows for the fainter GRBs. Finally, the
fraction of bright Swift GRB with redshift is only 29\% while that of
bright BeppoSAX and HETE2 GRBs is 62 \% and 50 \% respectively,
despite the much quicker optical follow-up observations for Swift
GRBs.  Excluding the objects with redshift obtained from host galaxy
emission lines from these samples does not change this conclusion.

Figure \ref{nhz}b) plots the redshift as a function of the observer
frame N$_H$ for the Swift and BeppoSAX GRB samples. Not surprisingly
the highest redshift GRBs are found not only at low peak-fluxes
(figure \ref{nhz}a), but also at low observed column densities. The
two dashed lines in figure \ref{nhz}b) are the expectation for a
constant, rest-frame column density of log$N_H=22$ (left line) and
log$N_H=23$ (right line). The observed Swift logN$_H$ distribution is
consistent with the expectation of rest frame column densities of the
order of 10$^{22}$ cm $^{-2}$, typical of dense molecular clouds. GRBs
with rest-frame obscuring column densities of the order of $10^{23}$
cm$^{-2}$ do exist. Such high column densities have been detected only
in high z GRBs so far (GRB050904 at z=6.29 and and GRB060510B at
z=4.9). These column densities imply a huge extinction of the
rest-frame UV light, if dust with properties similar to that in the
Galaxy, the SMC or even for a dust with a grain distribution strongly
shifted toward large grain sizes (Stratta et al. 2004, 2005) would be
associated to the X-ray absorbing gas. The simple detection of the
bright optical and near infrared afterglow of this GRB (Tagliaferri et
al.  2005, Haislip et al. 2006, Boer et al. 2006) implies peculiar
dust properties (Campana et al. 2006b, Stratta et al. 2007). 
Here we limit ourselves to note that high--z GRBs with a
gas column density similar to that of GRB050904 but with less extreme
dust properties would easily remain undetected in the optical and near
infrared.  Furthermore, their host galaxies would be so faint that
unambiguous associations with the GRB would be impossible, because the
probability to find such faint galaxies in the arcsec Swift XRT
error-boxes would be not negligible, thus making impossible the
determination of their redshift.

To assess more quantitatively how the different selection effects
(peak-flux limit, GRB obscuration and magnitude of the optical
afterglow) can modify the redshift distribution we extracted from the
Swift GRB sample a subsample having the same peak-flux, N$_H$ and
Rmag(at 10ks) distributions of the joined BeppoSAX+HETE2 sample (the
``constrained'' GRB sample hereafter).  Figure 7
compares the redshift distribution of the constrained GRB sample with
that of the full Swift and BeppoSAX+HETE2 GRB samples. To evaluate the
uncertainty on the constrained GRB sample redshift distribution we ran
the random extraction 100 times and plot the contours of the region
covered by the constrained GRB sample redshift distributions.  We see
that the constrained GRB sample redshift distribution is consistent,
to within the uncertainties, with the real BeppoSAX+HETE2 redshift
distribution.

Other, more subtle, selection effects may be at work as well. For
example, there are redshift ranges for which the typical interval
covered by optical spectrometers ($\approx$3800-8000 \AA\ ) does not
contain any strong emission or absorption line. For example, strong
emission lines such as H$\alpha ,~H\beta ,
~[OIII]\lambda\lambda4959,5007,~[OII]\lambda3725$ go out of the above
wavelength range at z$\sim1.1$, while Lyman-$\alpha$ enter the range
at z$\sim2.1$. The redshift range 1.1-2.1 is the so called ``redshift
desert''. Analogously, the strongest absorption feature, after
Lyman-$\alpha$ is the MgII$\lambda\lambda2796,2803$ doublet. This goes
in a region strongly affected by telluric features already at
z$\gs1.5$. So redshift determinations through absorption lines in low
signal to noise spectra are difficult in the redshift range
1.5-2.1. In any case, treating quantitatively these effects is
difficult, because of the very diverse quality of the optical
spectra of GRB afterglows. Unfortunately, because of the highly
variable nature of these events, afterglow observations have often 
performed in non-optimal conditions and instrument set-ups, and
most importantly, they cannot be repeated.

\section{Conclusions}

We have compared three well defined samples of long GRBs observed and
localized by Swift (122 GRBs), BeppoSAX (39 GRBs) and HETE2 (44 GRBs),
for a total of 205 objects. Secure spectroscopic redshifts have been
measured for 67 of these GRBs. The fraction of redshift determinations
is similar in the three samples, 34 \%, 30\% and 32\% respectively.

Swift GRBs are, on average, slightly fainter and harder than BeppoSAX
and HETE2 GRBs.  This is probably due to both the better sensitivity
of the BAT detector with respect to the BeppoSAX and HETE2 detectors
and to the higher energy range (15-150 keV) where Swift GRBs are
detected and localized, compared to BeppoSAX and HETE2 ($\approx2-20$
keV).  The distribution of the observer frame N$_H$ for the Swift GRBs
is shifted toward higher N$_H$ values than BeppoSAX, at a confidence
level of better than 98\%. This is again probably due to the different
energy bands in which GRBs are localized by the two satellites. The
most obscured GRBs have probably been missed by the BeppSAX survey.
The distribution of the observer frame N$_H$ for the Swift GRBs
without redshift determination is also shifted toward higher N$_H$
values than that of the Swift GRBs with a redshift determination
(confidence level of better than 99\%), implying that the sample of
Swift GRBs with redshift determinations is biased against large
obscuration. This is confirmed by a more detailed analysis of the
sample of bright GRBs. If dust is associated to the X-ray absorbing
gas, one would expect that extinction makes the discovery and study of
optical afterglows of bright Swift GRBs more difficult. This is
probably the case, since the fraction of bright Swift GRB with
redshift is only 29\% while that of bright BeppoSAX and HETE2 GRBs is
62 \% and 50 \% respectively.  Highly obscured, bright, low redhisft
GRBs are likely present in the Swift sample, but so far most of them
must have escaped redshift determination (we expect that the majority
of the 17 bright Swift GRBs without redshift are at z$\ls2$). A
program to discover and measure the magnitude and the redshift of the
host galaxies of bright Swift GRBs could confirm this conclusion and
provide a sample of GRB redshifts unbiased against obscuration.

\begin{figure}
\begin{center}
\includegraphics[height=8.5truecm,width=8.5truecm]{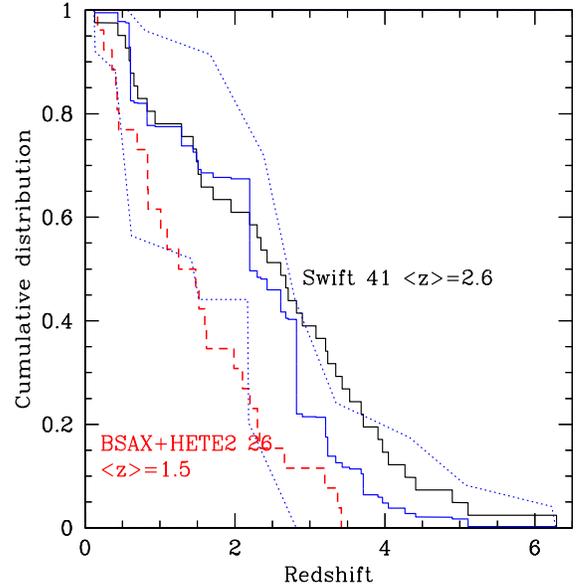}
\caption{The average cumulative redshift distribution of a subsample
of Swift GRBs having the same peak-flux, N$_H$ and Rmag(at 10ks)
distributions of the joined BeppoSAX+HETE2 samples (thin solid line)
compared with the Swift (thick solid line) and BeppoSAX+HETE2 (dashed
line) total redshift distributions. The thin dotted lines mark the
redshift range covered by 100 random extractions and give an idea
of the statistical uncertainty associated to a single extraction of a
redshift distribution of 26 GRBs from a parent population.}
\end{center}
\label{zdistsim}
\end{figure}

Highly X-ray obscured GRBs do exist also at high redshift. The
detection of bright optical and near infrared (UV rest frame)
afterglows from these GRBs implies a dust to gas ratio and/or dust
composition different from those of nearby GRBs (Stratta et
al. 2007). Indeed, at z$\gs5$ the major source of dust in the local
Universe (AGB stars) falls short of time to produce enough dust,
implying that high--z GRB host galaxies probably contains much less
dust than lower redshift host galaxies. This implies that redshift
determination of high--z GRBs would not be more difficult than that of
lower redshift GRBs, even if the observed optical and near infrared
bands sample the UV rest frame. If this is the case, the Swift sample
of GRBs with redshifts would be a fair sample of the real high--z GRB
population.

The absence of high redshift GRBs in the BeppoSAX and HETE2 samples of
GRBs with measured redshift is most likely due to the fact that
the median delay between the GRB event and the optical and near
infrared follow-ups for BeppoSAX and HETE2 GRBs is $\sim50$ times and
$\sim15$ times longer than that of Swift GRBs.  At the time of
BeppoSAX and HETE2 follow-up faint afterglow of high redshift GRBs are
too faint to allow redshift determination throght absorption line
spectroscopy.  Furthermore, the host galaxies of high redshift GRBs
are too faint to allow redshift determination through their emission
lines.  High redshift GRBs may well be present in the BeppoSAX and
HETE2 samples, but it is extremely difficult, if not impossible, to
determine their redshift and therefore recognize them as such.

Swift optical afterglows, measured at a fixed observer frame time,
e.g. 10 ks after the GRB event, are fainter than BeppoSAX and HETE2
optical afterglows, also when compared to the GRB 15-150 keV
peak-flux.  This is somewhat surprising, because the higher median
redshift of Swift GRBs implies that a fixed observer-frame time
samples, on average, a shorter rest-frame time delay from the GRB
event for the Swift GRBs than BeppoSAX and HETE2. Because afterglows
decrease like power laws one would expect that the ratio between the
GRB peak-flux and the optical afterglow magnitude at a fixed observed
time would be smaller for the Swift afterglows, contrary to what is
observed. At least two effects may contribute to explain the observed
trend. The first is that at z$>4$ the Lyman-$\alpha$ forest enters the
R band, thus reducing the observed optical flux. The second is a
higher extinction in Swift GRBs with respect to BeppoSAX and HETE2
GRBs, as disccused above.


To conclude at least selection effects on GRB localization and GRB
redshift determination must be properly taken into account in order to
safely use GRBs as cosmological tools, and derive the physical and
cosmological evolution of the GRB formation rate from statistical
analysis of the present GRB samples, at least selection effects on GRB
detection,. This would allow a fair and quantitively-meaningful
comparison with the star-formation rate estimated through other means.
Moreover, star-formation in regions hardly reachable by other
techniques (low mass, dwarf galaxies, high redshift galaxies, dust
enshshouded star-formation sites) could be probed.

{\bf Acknowledgments} We thank Rosalba Perna and Elena Rossi for early
discussions on the topics presented in this papers.  We also thank Eli
Waxman for useful comments and Luigi Stella for a careful reading of
the manuscript.  We acknowledge support from contracts ASI/I/R/039/04
and ASI/I/R/023/05/0.

\end{document}